\UseRawInputEncoding
\documentclass[aps,preprint,amsmath,amssymb,amsfonts,nofootinbib]{revtex4-1}
\usepackage{epsfig}
\usepackage{graphicx}
\usepackage{dcolumn}
\usepackage{bm}
\usepackage{amsthm}
\usepackage{amsmath}
\usepackage{color}

\newcommand{\beq}{\begin{equation}}
\newcommand{\eeq}{\end{equation}}
\newcommand{\bea}{\begin{eqnarray}}
\newcommand{\eea}{\end{eqnarray}}

\begin{document}

\title{A gauge theory for the 2+1 dimensional incompressible Euler equations}
\author{Christopher Eling}
\email{cteling@gmail.com}
\noaffiliation

\begin{abstract}

We show that in two dimensions the incompressible Euler equations can be expressed in terms of an abelian gauge theory with a Chern-Simons term.
The magnetic field corresponds to fluid vorticity and the electric field is the product of the vorticity and the gradient of the stream function.  This picture can be extended to active scalar models, including the surface quasi-geostrophic equation.  We examine the theory in the presence of a boundary and show that the Noether charge algebra is a Kac-Moody algebra.  We argue that this symmetry is associated with the nodal lines of zero magnetic field.

\end{abstract}

\maketitle

\section{Introduction}

In recent years a variety of connections have emerged between the properties of waves in fluid systems and the topological phases of matter \cite{Delplace, Souslov, Tauber1, Perrot, Tauber2, Grad, Liu:2020ksx,Liu:2020abb, Venaille, Green:2020uye, Tauber3}.  On the fluid dynamics side, the paradigmatic example is the set of 2+1 dimensional equations known as the shallow water equations.  These equations govern the behavior of a thin layer of fluid whose height fluctuations are much smaller than its horizontal extent.  For the height $h(x,y,t)$ and horizontal velocity $v^i(x,y,t)$, ($i = (x,y)$) the shallow water equations are
\begin{align}
\partial_t h + v^j \partial_j h + h \partial_j v^j = 0 \label{swconserv} \\
\partial_t v^i + v^j \partial_j v^i = f \epsilon^{i}_{j} v^j + g \partial^i h,
\end{align}
where $g$ is the gravitational acceleration and $f$ is the Coriolis parameter.  In the 19th century Lord Kelvin showed that the linearized version of these equations admits waves that are exponentially localized near the coasts.  These waves are chiral modes because they propagate only in one direction around the coast.  For example, for landmasses in the Northern (Southern) hemisphere the direction of the wave propagation is  clockwise (counterclockwise).  Later, it was found that chiral modes are also present at the equator, propagating only from east to west \cite{Matsuno, Higgins}.

Chiral edge modes also famously exist in topological phases of matter, particularly at the edge of quantum Hall systems \cite{Zhang,Wen}.  Quantum Hall fluids are described by an effective 2+1 dimensional gauge field theory with a Chern-Simons term.  For gauge connection $A_\mu = (A_0, A_i)$  this term has the form\footnote{Here $x^\mu = (t,x^i)$, but note that the expressions in this notation are not Lorentz covariant as the physics is non-relativistic.}
\begin{align}
S_{\rm CS} = \int d^3 x ~ \epsilon^{\mu \nu \rho} A_\mu \partial_\nu A_\rho \label{CSterm}.
\end{align}
The Chern-Simons term is invariant under gauge transformations $A_\mu \rightarrow A_\mu + \partial_\mu \lambda$ up to a total derivative.  In addition, this term is only sensitive to the topology of the manifold because it is independent of the metric tensor.

In \cite{Tong:2022gpg} D. Tong showed that the shallow water equations follow from the action of a 2+1 dimensional gauge theory with a type of Chern-Simons term.  At the linear level, the theory reduces to  Maxwell-Chern-Simons theory, with edge modes that match the coastal Kelvin waves.  Additional work has shown that other properties of the shallow water system can be expressed in terms of the gauge theory description \cite{Sheikh-Jabbari:2023eba}. More recently, it has also been shown that Tong's analysis can be extended to include a class of compressible Euler equations \cite{Nastase:2023rou}.

A natural question is whether other fluid dynamical systems can be expressed in terms of a gauge theory.  In this paper we investigate whether the incompressible 2+1 dimensional Euler equation and its ``active scalar" generalizations have a description in terms of a gauge theory. The incompressible Euler equations have the form
\begin{align}
\partial_i v^i = 0 \\
\partial_t v^i + v^j \partial_j v^i + \partial^i p = 0, \label{Eulereqn}
\end{align}
where $p$ is the fluid pressure and the constant fluid density has been set to unity.   We find that the fluid vorticity $\theta$ and the gradient of the stream function $\psi$ ($v^i = \epsilon^{ij} \partial_j \psi$) can be mapped into the pseudoscalar magnetic field and the electric field vector, respectively, by identifying the 2+1 dimensional Bianchi identity with the ideal vorticity equation
\begin{align}
\partial_t \theta + \epsilon^{ij} \partial_j \psi \partial_i \theta = 0  \label{vorticityeqn}.
\end{align}
We show that a gauge invariant action with non-topological terms together with a Chern-Simons term reproduces the Euler equation (\ref{Eulereqn}) and the correct relationship between $\psi$ and $\theta$ involving the 2d Laplacian
\begin{align}
\theta = -\Box \psi.
\end{align}
We also demonstrate that this action can be rewritten as a pure Chern-Simons action via a non-linear field redefinition.

Finally, generalizing these actions to include a non-local term reproduces the equations of active scalar models.  In these models, which include the surface quasi-geostrophic equation \cite{Held}, the relation between $\theta$ and $\psi$ is via a generalized fractional Laplacian operator.

The structure of this paper is as follows.  In Section \ref{gaugetheory} we review Tong's construction of the gauge theory for the shallow water equation.  We then use similar principles to construct a different gauge theory for the incompressible Euler and active scalar models. In Section \ref{edge} we discuss the presence of edge modes when our theory contains a boundary.  We show that the algebra of conserved Noether surface charges is a Kac-Moody algebra, which suggests the presence of conformal symmetry associated with the boundary.  We also argue that nodal lines of zero magnetic field (zero vorticity) are natural boundary surfaces in flows, and conjecture this explains the appearance of nodal lines with conformally invariant measures in the turbulent inverse cascade \cite{conformal1, conformal2}.  We conclude with a discussion of open questions, including the possibility of extending the construction to 3+1 dimensions.

\section{Constructing the gauge theory action}
\label{gaugetheory}

\subsection{Shallow water equations}

First, we review the construction of the gauge theory action for the shallow water equations.  To construct his action principle, Tong noted that the first shallow water equation (\ref{swconserv}) has the form of a conservation law $\partial_\mu J^\mu = 0$ with conserved current
\begin{align}
J^0 &= h \nonumber \\
J^i &= h v^i.
\end{align}
In a 2+1 dimensional gauge theory, the Bianchi identity implies
\begin{align}
\epsilon^{\mu \nu \rho} \nabla_\mu F_{\nu \rho} = \epsilon^{\mu \nu \rho} \partial_\mu \partial_\nu A_\rho = 0,
\end{align}
which can be rearranged into the form of the conservation of a current
\begin{align}
\partial_\mu \left(\epsilon^{\mu \nu \rho} F_{\nu \rho} \right) = 0.
\end{align}
In terms of the magnetic field $B = \epsilon^{ij} \partial_i A_j$ and electric field $E_i = \partial_t A_i - \partial_i A_0$,
\begin{align}
J^0 &= B \nonumber \\
J^i &= -\epsilon^{ij} E_j \label{Bianchicurrent}.
\end{align}
Identifying this topological current with the conserved current of the shallow water equation yields
\begin{align}
B &= h \\
E_i &= h \epsilon_{ij} v^j.
\end{align}
The next step is to postulate a Lagrangian in the form of the shallow water system's kinetic energy minus its potential energy:
\begin{align}
S_{\rm base} =  \int dt d^2 x \left(\frac{1}{2} h v^2 - \frac{1}{2} gh^2 \right)  = \int dt d^2 x \left(\frac{E^2}{2B}  - \frac{1}{2} gB^2 \right).
\end{align}
Varying with respect to $A_0$ gives the Gauss law
\begin{align}
\partial_i \left(\frac{E^i}{B} \right) = \theta = 0,
\end{align}
where $\theta = \epsilon^{ij} \partial_i v_j$ is the fluid vorticity.  Variation with respect to $A_i$ yields the dynamical equation
\begin{align}
\partial_t \left(\frac{E^i}{B} \right) + \frac{1}{2} \epsilon^{ij} \partial_j \left(\frac{E^i}{B}\right) + g \epsilon^{ij} \partial_j B = 0.
\end{align}
In terms of fluid variables
\begin{align}
\partial_t v^i + \frac{1}{2} \partial^i (v^2) + g \partial^i h = 0.
\end{align}
Clearly, the Lagrangian yields equations of motion that are almost consistent with the shallow water equations but constrain the fluid's vorticity to be zero (the Coriolis parameter $f$ is also neglected here).

In fact, taking the curl of the second shallow water equation yields a second conserved current associated with the vorticity, with
\begin{align}
\tilde{J}^0 &= \theta + f \nonumber \\
\tilde{J}^i &= (\theta + f) v^i
\end{align}
The idea is to account for non-zero vorticity by coupling the vorticity current to the gauge connection $A_\mu$.  One postulates an additional term in the action
\begin{align}
S_{\rm coupling} = \int dt d^2 x \left(\tilde{J}^0 A_0 + \tilde{J}^i A_i \right) = \int dt d^2 x~ \tilde{J}^\mu A_\mu.
\end{align}
This term is invariant under gauge transformations $A_\mu \rightarrow A_\mu + \partial_\mu \lambda$ because the vorticity current is conserved.  The vorticity current is then expressed in terms of Clebsch scalars $\alpha$ and $\beta$
\begin{align}
\tilde{J}^\mu = \epsilon^{\mu \nu \rho} \partial_\nu \beta \partial_\rho \alpha.
\end{align}
Varying the total action
\begin{align}
S_{\rm total,sw} = \int dt d^2 x \left(\frac{E^2}{2B}  - \frac{1}{2} gB^2 + f A_0 - \epsilon^{\mu \nu \rho} A_\mu \partial_\nu \beta \partial_\rho \alpha \right)
\end{align}
with respect to $A_0$, $A_i$, $\alpha$, and $\beta$ yields the shallow water equations (an additional $f A_0$ term must be included to account for the Coriolis parameter).  The final term has a Chern-Simons form if an additional auxiliary gauge connection $\tilde{A}_\mu$ is introduced:
\begin{align}
\tilde{A}_\mu = \partial_\mu \chi + \beta \partial_\mu \alpha \\
\epsilon^{\mu \nu \rho} A_\mu \partial_\nu \beta \partial_\rho \alpha =  \epsilon^{\mu \nu \rho} A_\mu \partial_\nu \tilde{A}_\rho.
\end{align}

\subsection{2d incompressible Euler equation}

We now follow a similar type of procedure for the 2+1 dimensional incompressible Euler equations.  To start, we note that the 2d incompressible Euler equation also has a conserved vorticity current.  The ideal vorticity equation has the form
\begin{align}
\partial_t \theta + v^j \partial_j \theta = 0.
\end{align}
When this equation is expressed in terms of the stream function $\psi$ we find Eqn. (\ref{vorticityeqn}) above.  This equation in turn can be expressed as the conservation of a current with components
\begin{align}
J^0 &= \theta \nonumber \\
J^i &= \theta \epsilon^{ij} \partial_j \psi.
\end{align}
Now we want to match these components to the identically conserved current associated with the 2+1 dimensional Bianchi identity.  Using Eqn. (\ref{Bianchicurrent}) above we find the following relationship between the fluid variables and the magnetic and electric fields
\begin{align}
B &= \theta \nonumber \\
E_i &= -\theta \partial_i \psi  \label{EBident}.
\end{align}
The next step is to postulate a gauge theory action.  Here we simply use the first term of Tong's action
\begin{align}
S_{\rm base} = \int dt d^2 x \left(\frac{E^2}{2B}\right).
\end{align}
As we saw earlier, variation of this action with respect to $A_0$ gives a Gauss law. In terms of the Euler equation fluid variables
\begin{align}
\partial_i \left(\frac{E^i}{B} \right) = -\Box \psi = 0.
\end{align}
This seems to imply that vorticity is zero as before.  However, at this stage there is no relationship between $\theta$ and $\psi$, they are independent variables in the action.  We can create a relationship by coupling the conserved vorticity current (i.e., the current from the Bianchi identity) to the gauge connection of our theory.  In this case there is no need to introduce Clebsch variables and the auxiliary gauge field.  The resulting coupling term is an explicit Chern-Simons term
\begin{align}
S_{\rm coupling} = \frac{1}{2} \int dt d^2 x~ J^\mu A_\mu = \frac{1}{2} \int dt d^2 x~ \left(B A_0 - \epsilon^{ij} E_j A_i \right) = \frac{1}{2} \int dt d^2 x ~\epsilon^{\mu \nu \rho} A_\mu \partial_\nu A_\rho.
\end{align}
Varying the action
\begin{align}
S = \int dt d^2 x \left (\frac{E^2}{2B} - \frac{1}{2} \epsilon^{\mu \nu \rho} A_\mu \partial_\nu A_\rho \right)
\end{align}
with respect to $A_0$ now yields the correct relationship between $\psi$ and $\theta$
\begin{align}
\partial_i \left(\frac{E^i}{B} \right) - B = 0 ~~ \Rightarrow  ~~  -\Box \psi = \theta.
\end{align}
The remaining step is to consider the variation with respect to $A_i$.  This gives
\begin{align}
-\partial_t \left(\frac{E^i}{B} \right) - \epsilon^{ij} \partial_j \left( \frac{E^2}{2B^2} \right) + \epsilon^{ij} E_j = 0.
\end{align}
Multiplying by $\epsilon_{ki}$ yields
\begin{align}
-\partial_t \left(\frac{\epsilon_{ki} E^i}{B} \right) + \partial_k \left(\frac{E^2}{2B} \right) - E_k = 0.
\end{align}
Note that
\begin{align}
\frac{\epsilon^{ij} E_j}{B} =  -\epsilon^{ij} \partial_j \psi = -v^i.
\end{align}
Therefore, the equation becomes
\begin{align}
\partial_t v_k + \frac{1}{2} \partial_k (v^2)  - \theta \epsilon_{kj} v^j = 0.
\end{align}
Using the identity
\begin{align}
\frac{1}{2} \partial_k (v^2)  = v^j \partial_j v^k +  \theta \epsilon_{kj} v^j,
\end{align}
produces the Euler equation (\ref{Eulereqn}) without a pressure term.   However, a pressure term follows simply by including a term $pB$ in the action, where $p$ is an auxiliary function.   This additional term does not change the Gauss law relation but does add the correct pressure gradient.  Thus, we arrive at the following action for the Euler system
\begin{align}
S_{\rm gauge,Euler} = \int dt d^2 x \left (\frac{E^2}{2B} - p B - \frac{1}{2} \epsilon^{\mu \nu \rho} A_\mu \partial_\nu A_\rho \right) \label{totalaction}.
\end{align}
To summarize, the ideal 2d vorticity equation follows from the Bianchi identity of the gauge theory.  The Gauss law constraint relates the stream function to the vorticity, and the dynamical equation of motion is the Euler equation.

Although this is a type of Chern-Simons action, the theory is different from standard Maxwell-Chern-Simons theory.  For example, in Maxwell-Chern-Simons theory the Maxwell parts (i.e., $E^2-B^2$) of the action are invariant (even) under time reversal, while the Chern-Simons term changes sign (odd).  Hence, standard Maxwell-Chern-Simons theory is not time reversal invariant.

In our case, we assume that when $t \rightarrow -t$, $x^i \rightarrow x^i$, $A_0 \rightarrow A_0$ and $A_i \rightarrow -A_i$.  Under these transformations, the electric field $E_i$ is even, but the magnetic field is odd: $B \rightarrow -B$.  This is consistent with fluid dynamics, where the vorticity changes sign under time reversal, but the combination of velocity and vorticity (i.e., the electric field) is even.

All three terms in the action (\ref{totalaction}) are odd, and therefore the total action is odd under time reversal.  Hence, the resulting classical equations of the motion are time reversal invariant, as expected. Note however, in a statistical mechanics picture, the partition function $Z = \int [dA] \exp(-\beta S_{\rm gauge})$ would apparently change under time reversal.

\subsection{Non-linear field redefinition}

In \cite{Lemes:1997vx} it was shown that any Yang-Mills type of action in the presence of a Chern-Simons term (i.e., topologically massive Yang-Mills theory) can be reabsorbed into a pure Chern-Simons theory via a non-linear, but local field redefinition.  In the simplest abelian case
\begin{align}
S_{\rm Maxwell}(A)+ S_{\rm CS}(A) = S_{\rm CS}(\hat{A}),
\end{align}
where
\begin{align}
\hat{A}_\mu = A_\mu +  \sum_{n=0}^{\infty} \frac{1}{m^n} C^n_\mu(\partial,F).
\end{align}
The terms $C_\mu$ depend on the field strength $F$ and its derivatives.  $m$ is a mass parameter that accounts for the difference in dimensions between the Maxwell action, which is second order in derivatives and the Chern-Simons action, which is first order in derivatives.

In our theory, both non-topological terms are first order in derivatives, just like the Chern-Simons term.  This suggests that these terms may be reabsorbed into a pure Chern-Simons term via simpler field redefinition. We propose the following redefinition of the gauge potential $A_\mu = (A_0,A_i)$
\begin{align}
\hat{A}_0 &= A_0 + p - \frac{E^2}{2B^2} \nonumber \\
\hat{A}_i &= A_i.
\end{align}
Substituting this field redefinition into the pure Chern-Simons action constructed from $\hat{A}_\mu$ reproduces the action (\ref{totalaction}) up to total derivatives
\begin{align}
-\frac{1}{2} \int d^3 x ~ CS(\hat{A}) = \int d^3 x \left(- \frac{1}{2} CS(A) - pB + \frac{E^2}{2B} \right),
\end{align}
where $CS$ denotes the Chern-Simons term in (\ref{CSterm}). The variation of the Chern-Simons action with respect to $\hat{A}_\mu$ yields the field equations, which impose the vanishing of the magnetic and electric fields of the hatted gauge potential
\begin{align}
\hat{B} &= B = 0 \nonumber \\
\hat{E}_i &= E_i - \partial_i p + \frac{1}{2} \partial_i \left(\frac{E^2}{B^2}\right) = 0.
\end{align}
In fluid language these equations correspond to those satisfied by steady, irrotational flows.  Hence, the space of steady, irrotational solutions on a manifold corresponds to the space of flat connections on the manifold.

\subsection{Active scalar models}

The action that generates the Euler fluid system described above can also be generalized to a produce the wider set of ``active scalar'' models.  In these cases, the conservation equation has the same form as before
\begin{align}
\partial_t \theta + \epsilon^{ij} \partial_j \psi \partial_i \theta = 0.
\end{align}
However, the relationship between $\theta$ and the stream function $\psi$ takes the following form in terms of a fractional Laplacian
\begin{align}
-\Box^{m/2} \psi = \theta.
\end{align}
Equivalently,
\begin{align}
\psi(\vec{x}, t) =  \int d^2 \vec{y} ~  |\vec{x} - \vec{y}|^{m-2} ~ \theta(\vec{y}, t).
\end{align}
Different physical models are described by different values of the parameter $m$.  $m=2$ is the Euler system.  Other values of physical interest are $m=1$, which is the surface quasi-geostrophic (SQG) model \cite{Held}.  In this case $\theta$ is a ``potential temperature".  The case $m=-2$ describes large scale flows of a rotating shallow fluid (known as the Charney-Obhukov-Hasegawa-Mima model) \cite{McWilliams} in the limit of vanishing Rossby radius.

The associated generalized Euler equation has the form \cite{Tao}
\begin{align}
\partial_t ({\cal A} v^i) + v^j \partial_j ({\cal A} v^i) + ({\cal A} v^j) \partial_k v_j + \partial^i p = 0,
\end{align}
where the operator ${\cal A} = \Box^{m/2-1}$.  In terms of the electric and magnetic field identifications (\ref{EBident}) the Gauss law and dynamical equation are
\begin{align}
\partial_i \Big( {\cal A} \left(\frac{E^i}{B} \right) \Big) &= B  \\
-\partial_t \Big( {\cal A} \left( \epsilon_{ki} \frac{E^i}{B} \right) \Big) + \partial_k \Big( \frac{E_i}{B}~ {\cal A} \left(\frac{E^i}{B} \right) \Big) - E_k + \partial_k p &= 0.
\end{align}
These equations follow from replacing the first term in (\ref{totalaction}) with the following nonlocal term, which formally has the form
\begin{align}
L_{\rm nonlocal} = {\cal A} \left(\frac{E^i E_i}{2B} \right).
\end{align}
For $0<m<2$ this term in the action can be written as
\begin{align}
L_{\rm nonlocal} \sim \int d^2 \xi  \frac{1}{|\vec{\xi}|^m}  \left(\frac{E^i(\vec{x}-\vec{\xi}) E_i(\vec{x}-\vec{\xi})}{2 B(\vec{x}-\vec{\xi})}\right).
\end{align}
For $2<m<4$, the non-local term is expressed as a singular integral
\begin{align}
L_{\rm nonlocal} \sim \int d^2 \xi  \frac{1}{|\vec{x}-\vec{\xi}|^m} \left( \frac{E^i(\vec{x}) E_i(\vec{x})}{2 B(\vec{x})} - \frac{E^i(\vec{\xi}) E_i(\vec{\xi})}{2 B(\vec{\xi})} \right).
\end{align}

\section{Edge Modes and Symmetries}
\label{edge}

The presence of the explicit Chern-Simons term in the action has several interesting consequences when the theory is considered in the presence of a boundary.   In particular, the interplay of boundary conditions and gauge invariance lead to the existence of edge modes, which was first discovered in the context of the quantum Hall effect (see, e.g., \cite{Wen:1992vi, Tong:2016kpv}).

The starting point is the observation that under a gauge transformation $A_\mu \rightarrow A_\mu + \partial_\mu \lambda$ the Lagrangian (\ref{totalaction}) is only invariant up to a total derivative
\begin{align}
\delta_\lambda S_{\rm total} = -\frac{1}{2} \int dt d^2 x ~ \partial_\mu (\lambda J^\mu),
\end{align}
which for a spatial boundary leads to a extra boundary correction
\begin{align}
\delta_\lambda S_{\rm total} = -\frac{1}{2} \int dx  \lambda ~ (J^\mu n_\mu).
\end{align}
Using the fluid identifications (\ref{EBident}) this term becomes
\begin{align}
\delta_\lambda S_{\rm total} = -\frac{1}{2} \int  dx  \lambda (E_i t^i)  = -\frac{1}{2} \int dx ~ \lambda (v^i n_i),
\end{align}
for tangent vector $t^i$ and normal $n^i$.   To fix this problem, one approach is to restrict the gauge parameter $\lambda$ to vanish on the boundary.  However,  in fluid dynamics one natural choice of boundary conditions is to restrict the flow to be purely tangential to the boundary, in which case the anomalous boundary term is zero.

This is not the end of the story because an action principle only correctly produces the equations of motion when boundary terms are set to zero.  For the variation of (\ref{totalaction}) the boundary terms are
\begin{align}
\delta S_{\rm bdry} = \int dt dx \left( -\frac{n_i E^i}{B}~ \delta A_0 + \frac{1}{2} \epsilon^{ij} n_i A_j~ \delta A_0 - \frac{E^2}{2B^2} \epsilon^{ij} n_i \delta A_j - p \epsilon^{ij} n_i \delta A_j - \frac{1}{2} A_0 \epsilon^{ij} n_i \delta A_j  \right).
\end{align}
One way these boundary terms can be set to zero is if $A_0$ and the component of $A_i$ parallel to the boundary are fixed to be constants on the boundary.  Note that this restricts the allowed gauge transformations to the subset of transformations that preserve these boundary conditions.  Typically, local gauge transformations are redundancies in the description of the theory and are distinct from global symmetries acting on actual physical degrees of freedom.  However, restricting the gauge freedom allows formerly unphysical degrees of freedom to become physical.  Hence the restricted gauge transformations preserving the boundary become global symmetries of the boundary degrees of freedom.

The conserved charges associated with these symmetries are given by the Noether charges associated with the gauge transformations.  Schematically, the variation of any Lagrangian yields
\begin{align}
\delta L = E_\mu \delta \rho^\mu + \partial_\mu \theta^\mu,
\end{align}
where $E_\mu$ are the equations of motion, $\rho^\mu$ represents the fields, and $\theta^\mu$ is the symplectic current.  Furthermore, the Lagrangian itself is a total derivative under the transformation $\delta L = \partial_\mu S^\mu$.  Consequently, the Noether current $N^\mu = S^\mu-\theta^\mu $ is conserved on-shell when the equations of motion are imposed
\begin{align}
\partial_\mu N^\mu \approx 0
\end{align}
On-shell $N^\mu$ can be expressed as the divergence of a superpotential 2-form $P^{\mu \nu}$, i.e $N^\mu \approx \partial_\nu P^{\mu \nu}$.  Hence, the Noether charge $Q$ of the theory is a surface charge via
\begin{align}
Q = \int N^\mu d\Sigma_\mu =  \int P^{\mu \nu} d\Sigma_{\mu \nu}.
\end{align}

For the theory (\ref{totalaction})  the resulting Noether current for gauge transformations ($\delta A_\mu = \partial_\mu \lambda$) is
\begin{align}
N^0 =& -\frac{E^i}{B} \partial_i \lambda - \frac{1}{2} \epsilon^{ik} A_i \partial_k \lambda + \frac{1}{2} B \lambda \nonumber \\
N^i =& \frac{E^i}{B} \partial_t \lambda - \frac{1}{2} \epsilon^{ij} A_j \partial_t \lambda + \frac{E^2}{2B^2} \epsilon^{ij} \partial_j \lambda + p \epsilon^{ij} \partial_j \lambda + \frac{1}{2} \epsilon^{ij} A_0 \partial_j \lambda - \frac{1}{2} \epsilon^{ij} E_j \lambda.
\end{align}
Using these results yields the following set of $\lambda$ dependent Noether charges
\begin{align}
Q_\lambda = \int dx \lambda \left(\frac{E^i}{B} - \frac{1}{2} \epsilon^{ij} A_j \right) n_i.
\end{align}
In terms of fluid variables, and assuming the boundary is closed surface, the charges have the form of a $\lambda$ dependent local circulation
\begin{align}
Q_\lambda = \oint \lambda \left(v^i - \frac{1}{2} A^i \right) dx_i.
\end{align}
A similar result was found in \cite{Sheikh-Jabbari:2023eba} for the Maxwell-Chern-Simons theory describing linearized shallow water waves.  The infinite number of charges indicates the presence of an infinite dimensional symmetry describing the boundary system.  The algebra of these charges follows from the Poisson bracket
\begin{align}
\{Q_{\lambda_1}, Q_{\lambda_2}\} = \delta_{\lambda_2} Q_{\lambda_1}.
\end{align}
One can define a ``charge aspect" $\gamma = \left(v^i - \frac{1}{2} A^i \right) dx_i$.  The charge aspect is not gauge invariant, changing by a total derivative $t^i \partial_i \lambda$.  Note that the global circulation charge (i.e. $Q_\lambda$ for $\lambda = 1$) is gauge invariant for singled valued gauge transformations.

The gauge non-invariance of $\gamma$ leads to a central extension term in the charge algebra, which is a U(1) affine Kac-Moody algebra in the form
\begin{align}
\{\gamma(t,x), \gamma(t,x')\} = -\frac{1}{2} \partial_x \delta(x-x').
\end{align}

The appearance of the Kac-Moody algebra is typically associated with a chiral boson theory on the boundary \cite{Elitzur:1989nr, Balachandran:1991dw,Park:1998yw}.  The simplest theory of a two-dimensional boson field $\Phi$ can be defined via the action
\begin{align}
S_{\rm boson} \sim \int d^2 x ~\partial_\mu \Phi \partial^\mu \Phi.
\end{align}
Solutions to this theory (harmonic functions) naturally separate into left and right movers, or equivalently in the complex plane, holomorphic and antiholomophic parts, i.e. $\Phi(z,\bar{z}) = \phi(z) + \bar{\phi}(\bar{z})$.  It is well known that the holomorphic (chiral) conserved current of the theory ${\cal J} = i \partial \phi$ obeys the U(1) Kac-Moody algebra (ie, a current algebra).  In terms of a mode expansion ${\cal J}  = \Sigma_n ~ {\cal J}_n z^{-n-1}$
\begin{align}
\{{\cal J}_n, {\cal J}_m\} = -\frac{1}{2} m \delta_{m+n,0}
\end{align}
Furthermore, the stress tensor $T$ of the theory can be constructed as the product of two currents $T =-(1/2) (\partial \phi)^2$.  The algebra associated with the stress tensor mode expansion is the famous Virasoro algebra associated with conformally invariant field theories \cite{Goddard:1986bp, Ginsparg:1988ui}.

The appearance of the Kac-Moody algebra indicates that there is an underlying conformal symmetry associated with boundaries in incompressible fluid flows.  A natural ``edge" is the boundary between regions of positive values and negative values of the magnetic field $B$, represented by nodal lines of $B$, which correspond to boundaries where $\theta = 0$. At these boundaries, the electric field also vanishes, but the fluid velocity defined as the ratio of $E^i$ and $B$ is finite.  Also, the Hamiltonian is zero, switching signs between regions of positive and negative values.

Topologically protected modes appear when time reversal symmetry is broken, as in the shallow water system with non-zero Coriolis parameter.  While the incompressible Euler equations are time reversal invariant, two-dimensional turbulence is asymmetric under time reversal.  Interestingly, in 2006 it was shown numerically that zero vorticity nodal isolines in the turbulent 2d inverse cascade are identical to the curves forming cluster boundaries in the scaling limit of statistical mechanical systems at equilibrium \cite{conformal1, conformal2}.  Turbulent flows in 2d can be created by stirring or exerting a random force on the fluid at some scale $L_f$.  Kraichnan \cite{Kraichnan} discovered that a ``double cascade" appears, where the enstrophy created by stirring is transferred to small scales (direct cascade), while energy is transferred to large scales (inverse cascade).  When a large separation of scales $L >> L_f$ exists, the physics of the inverse cascade is expected to be independent of the details of the forcing.  Moreover, unlike in 3d turbulence, the effects of viscosity at small scales can be neglected, and some properties of ideal flows are likely to be relevant.

The nodal curves are expected to random fractal lines in turbulence, but it is surprising that the measure on the curves is conformal and described by Schramm-Loewner evolution (SLE) \cite{Schramm}.  SLE curves have been proven to be the scaling limits of cluster boundaries formed by various statistical mechanical models defined on a lattice.  At criticality and in the continuum limit, these models are described by different 2d conformal field theories (for a review, \cite{Bauer:2006gr}).  Similarly, different fluid models (ie, different $m$) correspond to different 2d CFT's \cite{Falkovich} even though the turbulent inverse cascade system is far from equilibrium.

Because the edge states are associated with topology, it is likely that they are robust in highly turbulent flows at large Reynolds number.  In the turbulent state a statistical description of the flow is required and the Kac-Moody and Virasoro symmetries of the boundary edges should be associated with probabilities.  We conjecture that the probability that one or more $B=0$ curves is present in an infinitesimal region around a point in the inverse cascade is conformally invariant and is associated with the stress tensor operator of a chiral boson CFT at that point \cite{Eling}.

\section{Discussion}

In this paper we have shown that the incompressible Euler system in 2+1 dimensions can be described by a $U(1)$ gauge invariant action with an explicit Chern-Simons term.  Furthermore, we argued that this gauge theory picture also can describe the so-called active scalar models which include the surface quasi-geostropic equation.  In the presence of a boundary, a gauge theory includes edge states, which are physical degrees of freedom that live on the boundary surface.  We showed that the conserved Noether charge associated with closed boundaries in our gauge theory exhibits a Kac-Moody algebra, which indicates the presence of a chiral boson boundary theory with an underlying conformal symmetry.  Finally, we argued that zero magnetic field (i.e., zero vorticity) nodal lines are natural boundaries in flows and that fluctuations of these lines have a chiral boson description.  This may explain the mysterious appearance of conformal invariant measures of nodal lines and the SLE description of these lines in the turbulent inverse cascade.

In future, it would be interesting to explore whether additional properties of fluid systems can be encoded in gauge theory language and vice versa. For example, it would be interesting to connect the gauge theory picture with Arnold's idea of the Euler equation as a geodesic flow on the infinite dimensional Riemannian manifold of area preserving diffeomorphisms \cite{Arnold}.  One observation is that in two dimensions linearized area preserving diffeomorphisms in the space of the Lagrangian (co-moving) coordinates act as U(1) gauge transformations \cite{Susskind:2001fb, Sheikh-Jabbari:2023eba}.

Another interesting question is whether can one extend the action to include the fluid viscosity term, thereby giving a description of the full Navier-Stokes equation.  Because a viscous term modifies the vorticity equation (\ref{vorticityeqn}), non-zero viscosity should break the gauge invariance of the theory.  A related area of further work would be to investigate whether the gauge theory description can shed more light on the mechanism and properties of turbulence in 2d.

Finally, a natural area of investigation is whether there is a gauge theory description of the 3+1 incompressible Euler equation.  In this case vorticity is a vector $\omega^i$ and the ideal vorticity equation has the form
\begin{align}
\partial_t \omega^i + v^j \partial_j \omega^i - \omega^j \partial_j v^i = 0.
\end{align}
For a 3+1 dimensional gauge theory the Bianchi identity is conservation of a two-form
\begin{align}
\partial_\mu \left(\epsilon^{\mu \nu \rho \sigma} F_{\rho \sigma}\right) = 0.
\end{align}
These equations yield the familiar Maxwell equations
\begin{align}
\partial_i B^i  = 0 \nonumber \\
\partial_t B^i +  \epsilon^{ijk} \partial_j E_k = 0.
\end{align}
Hence, we can identify the magnetic and electric fields as
\begin{align}
B^i &= \omega^i \nonumber \\
E^i  = - \epsilon^{ijk} v_j \omega_k &=  -\vec{v} \times \vec{B}.
\end{align}
The fluid velocity can be expressed as
\begin{align}
\vec{v} = \frac{\vec{E} \times \vec{B}}{|B|^2} + \frac{H \vec{B}}{|B|^2}
\end{align}
where $H$ is the fluid helicity density $H=\vec{v}\cdot \vec{B}=\vec{v}\cdot \vec{\omega}$.

The next step, in principle, would be to develop an action with a constraint equation relating the curl of the velocity to the vorticity.  However, a major problem in the construction of the action principle is that there is no Chern-Simons term in 3+1 dimensions.  Therefore, a simple extension of our argument to one higher dimension apparently will not work.  Perhaps the mapping between fluids and gauge fields is a part of the magic of 2+1 dimensions, but it may be possible to construct a gauge theory description using different techniques.  If topological terms are a generic feature of the gauge theory description of fluids, then a putative action in $n>3$ spacetime dimensions may involve a ``BF" term, which the exterior product of a $n-2$ form $\mathbf{B}$ (not to be confused with the magnetic field vector) and the field strength $F$.

\end{document}